\documentclass{article}%
\usepackage{amsfonts}
\usepackage{amsmath}
\usepackage{geometry}
\usepackage{graphicx}
\usepackage{hyperref}
\usepackage{tabularx}
\usepackage{amssymb}%
\providecommand{\U}[1]{\protect\rule{.1in}{.1in}}

\begin{document}

\title{Exact Solutions of D-dimensional Klein-Gordon Oscillator with Snyder-de Sitter Algebra}
\author{{Zoubir Hemame}$^{1,2}$
\and {Mokhtar Falek}$^{2}$
\and {Mustafa Moumni}$^{2,a}$\\
$^{1}$Department of Matter Sciences, University of Khenchela, ALGERIA \\
$^{2}$Laboratory of Photonic Physics and Nano-Materials (LPPNNM)\\
Department of Matter Sciences, University of Biskra, ALGERIA\\
$^{a}$correspondant author m.moumni@univ-biskra.dz}
\maketitle

\begin{abstract}
We study the effects of Snyder-de Sitter commutation relations on relativistic bosons by solving analytically in the momentum space representation the Klein-Gordon oscillator in arbitrary dimensions. The exact bound states spectrum and the corresponding momentum space wave functions are obtained using Gegenbauer polynomials in one dimension space and Jacobi polynomials in D dimensions case. Finally, we study the thermodynamic properties of the system in the high temperature regime where we found that the corrections increase the free energy but decrease the energy, the entropy and the specific heat which is no longer constant. This work extends the part concerning the Klein-Gordon oscillator for the Snyder-de Sitter case studied in two-dimensional space in J. Math. Phys. 60, 013505 (2019).


\textbf{P. A. C. S.} 03.65.Ge, 03.65.Pm.
\end{abstract}

\tableofcontents

\section{Introduction}

The relativistic harmonic oscillator is one of the most used, among fundamental physics models, in experimental studies to explain confinement since it is characterized by bound states with non-zero residual energy. For example, in nuclear physics domain, RHO is the central potential of the nuclear shell model and it has been used also as the confining two-body potential for quarks in particle physics. Since the article of Moshinsky and Szczepaniak \cite{Moshinsky}, Dirac oscillator (DO) raised a considerable attention by many researchers and thus has been studied intensively \cite{Dixit, Martinez, Hoa, Wu, Ferkous}. In addition, the study of this model recorded an extension to boson cases such as the Klein-Gordon oscillator (KGO) for spin 0 bosons \cite{Bruce, Dvoeglazov} and Duffin-Kemmer-Petiau (DKPO) for spins 0 and 1 particles \cite{Nedjadi1, Nedjadi2, Nedjadi3, Kulikov}. We also find the KGO in arbitrary dimensions which is established in some series of articles \cite{Chargui, Garcia}.

When talking about dimensions other than three, the harmonic oscillator is even more interesting because of the correspondence that exists between it and the Coulomb potential found in 2D systems \cite{Fu, Tiwari} or between 4D harmonic oscillator and 3D Coulomb potential \cite{Meer, Kazakov} (and the references therein). It is also used for modeling Landau levels in topological insulators \cite{Li, Kotulla, Kurkcuoglu} and we mention here the recent realization of a 4D spinless topological insulator through the achievement of an explicit construction of a 2D electric circuit lattice \cite{Yu}; this opens the way to realistic and ideal platforms to create higher-dimensional topological states in the laboratory.

On the other hand, there have been many attempts to study deformed quantum mechanical systems as they have a significant impact on the absorption of infinities that lie in the standard field theories. This was initially proposed in the Snyder model which has suggested that measurement in noncommutative quantum mechanics can be governed by a generalized uncertainty principle (GUP) \cite{Snyder}. Therefore, the fundamental length scale is supposed to be of the order of the Planck length \cite{Kempf}, which in turn leads to a minimum uncertainty in the position measurement. This approach is motivated by several physics theories such as noncommutative geometries \cite{Douglas}, doubly special relativity (DSR) \cite{Amelino}, string theories \cite{Capozziello} and black hole physics \cite{Scardigli}.

It is also legitimate to seek to incorporate the other pillar of modern physics, namely the theory of gravitation, in this construction,  through its most striking manifestation which is curved space-time. Therefore, many attempts have been made to develop the equivalent of Snyder model in the study of quantum mechanics in curved space-time by looking for some generalizations of the Heisenberg algebra by adding small corrections to the commutation relations such us the GUP \cite{Mignemi} as well as the extended uncertainty principle (EUP) \cite{Ghosh}; this with the main aim of integrating the combined effects of the non-commutative geometry and the theory of gravity in quantum mechanics. Recently, certain relativistic and non-relativistic problems have been solved within this framework of curved Snyder model and we cite as an example: the Schr\"{o}dinger oscillator system, the non-relativistic free particle and the DO \cite{Mignemi1, Stetsko}.

The main purpose of this work is to investigate the deformed quantum formulation of the KGO in arbitrary dimensions in a deformed space obeying the Snyder-de Sitter (SdS) algebra with an emphasis on determination of the thermodynamic functions which play a significant role in understanding physical properties. We mention here the study of the thermodynamic properties of the ordinary DO which has received a great amount of attention for their description of the quark--gluon plasma model \cite{Moreno, Pacheco, Ravndal}, the deformed DO with a minimal length in a thermal bath \cite{Nouicer}, where it was shown that there is an upper bound for the minimal length, and the one-dimensional DKPO with the SdS model which has significant importance in heavy quark systems \cite{Falek}. We extend in this work the part concerning the KGO case in the SdS space studied in 2D in \cite{Falek19}.

The outline of this paper is as follows: In the following section \ref{sec:SDS}, we give a review on the SdS model. In the third section \ref{sec:1D}, we solve exactly the one-dimensional KG equation for the oscillator-like interaction with deformed SdS algebra in the momentum space representation. We obtain the exact wave function and energy spectrum for this system. We discuss also some special cases and the corresponding numerical results. We extend the same study to an arbitrary dimension in section four \ref{sec:ND} where, by a straightforward calculation, we deduce the normalized wave functions and the energy spectrum. In the regime of high temperatures, the thermodynamic properties of the system are investigated and discussed numerically in the fifth section \ref{sec:TP}. The concluding remarks are given in the last section \ref{sec:CC}.

\section{Review of the deformed quantum mechanics relation}
\label{sec:SDS}

In the non-relativistic SdS model, the deformed Heisenberg algebra in three-dimensional case is defined by the following commutation relation \cite{Mignemi1, Stetsko}:
\begin{equation}
\left[X_{i},P_{j}\right]=i\hbar\left(\delta_{ij}+\alpha_{1}X_{i}X_{j}+\alpha_{2}P_{i}P_{j} +\sqrt{\alpha_{1}\alpha_{2}}\left(X_{i}P_{j}+P_{i}X_{j}\right)\right)
\label{eqt1}
\end{equation}
\begin{equation}
\left[X_{i},X_{j}\right]=i\hbar\alpha_{2}\xi_{ijk}L_{k}\text{ , }\left[P_{i},P_{j}\right]=i\hbar\alpha_{1}\xi_{ijk}L_{k}
\label{eqt2}
\end{equation}
where $\alpha_{1}$ and $\alpha_{2}$ are small positive parameters defining the deformations coming from Snyder algebra and de Sitter space respectively and $L_{k}=\xi_{ijk}X_{i}P_{j}$ denotes the components of the angular momentum operator satisfying the usual algebra:
\begin{equation}
\left[L_{i},X_{j}\right]=i\hbar\xi_{ijk}X_{k}\text{ , }\left[L_{i},P_{j}\right]=i\hbar\xi_{ijk}P_{k}\text{ , }\left[L_{i},L_{j}\right] =i\hbar\xi_{ijk}L_{k}
\label{eqt3}
\end{equation}
In the same manner as in ordinary quantum mechanics, the commutation relation \ref{eqt1} gives rise to uncertainty Heisenberg relations ($\gamma _{ij}=\left( \sqrt{\alpha _{1}}\left( X_{i}\right) +\sqrt{\alpha _{2}}\left( P_{j}\right) \right) ^{2}\geqslant 0$):
\begin{equation}
\triangle X_{i}\triangle P_{j}\geq\frac{\hbar}{2}\left(\delta_{ij}+\gamma_{ij}+\alpha_{1}\left(\Delta X_{i}\right)^{2} +\alpha_{2}\left(\Delta P_{i}\right)^{2}-2\sqrt{\alpha_{1}\alpha_{2}}\Delta X_{i}\Delta P_{j}\right)\text{ , }i=1,2,3
\label{eqt4}
\end{equation}

For its part, relation \ref{eqt2} implies the appearance of a nonzero minimal length in position and momentum uncertainties:
\begin{equation}
\left( \Delta X\right) _{\min }=\hbar \sqrt{\frac{\alpha _{2}\left( 1+\gamma\right) }{1+2\sqrt{\alpha _{1}\alpha _{2}}}}\text{ , }\left( \Delta P\right)_{\min }=\hbar \sqrt{\frac{\alpha _{1}\left( 1+\gamma \right) }{1+2\sqrt{\alpha _{1}\alpha _{2}}}}
\label{eqt5}
\end{equation}
The noncommutative operators $X_{i}$ and $P_{i}$ satisfying the SdS algebra \ref{eqt1} correspond to the rescaled uncertainty relation \ref{eqt4} in position and momentum space. In order to study quantum mechanical problems, we represent these operators as functions of the operators $x_{i}$\ and $p_{i}$, satisfying the usual canonical commutation relations of the ordinary quantum mechanics. But because of the noncommutative relations \ref{eqt2}, there is no space or momentum representation. We write $X_{i}$ and $P_{i}$ according to $p_{i}$ and $\partial _{p_{i}}$ with the following transformations and we use the term \textquotedblleft momentum\textquotedblright representation for these transformations:
\begin{align}
X_{i}& =i\hbar \sqrt{1-\alpha _{2}p^{2}}\partial _{p_{i}}+\lambda \sqrt{\frac{\alpha _{2}}{\alpha _{1}}}\frac{p_{i}}{\sqrt{1-\alpha _{2}p^{2}}}
\label{eqt6} \\
P_{i}& =-i\hbar \sqrt{\frac{\alpha _{2}}{\alpha _{1}}}\sqrt{1-\alpha _{2}p^{2}}\partial _{p_{i}}+\left( 1-\lambda \right) \frac{p_{i}}{\sqrt{1-\alpha _{2}p^{2}}}
\label{eqt7}
\end{align}
where $p$ vary in the domain $\left] -1/\sqrt{\alpha _{2}},1/\sqrt{\alpha_{2}}\right[ $ and $\lambda $ is an arbitrary real constant.

\section{One-Dimensional Klein-Gordon Oscillator}

\label{sec:1D}

The stationary equation describing the KGO in one dimensional (1-dim) space is written \`{a} la Moshinsky \cite{Moshinsky}:
\begin{equation}
\left[ c^{2}\left( P+im\omega X\right) \left( P\mathbf{-}im\omega X\right)+m^{2}c^{4}-E^{2}\right] \psi \left( p\right) =0
\label{eqt8}
\end{equation}
where $m$ is the mass of the particle, $\omega $ is the classical oscillator frequency and $c$ is the speed of light.

Using the commutation relation \ref{eqt1} and the momentum space realization of the operators $X$ and $P$, we get:
\begin{equation}
\left[ m^{2}\omega ^{2}\left( 1-\frac{\hbar \alpha _{1}}{m\omega }\right) X^{2}+\left( 1\mathbf{-}m\omega \hbar \alpha _{2}\right) P^{2}-m\omega \hbar\sqrt{\alpha _{1}\alpha _{2}}\left( PX+XP\right) -\varepsilon \right] \psi\left( p\right) =0
\label{eqt9}
\end{equation}
with $\varepsilon =m\hbar \omega +\left( E^{2}-m^{2}c^{4}\right) /c^{2}$.

We use the definition of SdS algebra from \ref{eqt6} and \ref{eqt7}, we rewrite this equation in the deformed momentum space:
\begin{equation}
\left\{ \hbar ^{2}\frac{\alpha _{1}}{\alpha _{2}}\gamma \gamma ^{\ast}\left( \sqrt{1-\alpha _{2}p^{2}}\frac{\partial }{\partial p}\right) ^{2}-\frac{2i\hbar \Omega }{\alpha _{1}\alpha _{2}}p\frac{\partial }{\partial p}-\frac{\eta \alpha _{2}p^{2}}{1-\alpha _{2}p^{2}}+\epsilon \right\} \psi\left( p\right) =0
\label{eqt10}
\end{equation}
with:
\begin{align}
\gamma & =\left( 1+im\omega \sqrt{\frac{\alpha _{2}}{\alpha 1}}\right) \text{ , }\Omega =\alpha _{1}\left( \gamma \gamma ^{\ast }\lambda -1\right)  \notag\\
\eta & =\frac{1-\lambda }{\alpha _{2}}-m\omega \hbar +\left( \frac{\lambda }{\alpha _{1}\alpha _{2}}+\frac{i\hbar }{\sqrt{\alpha _{1}\alpha _{2}}}\right) \Omega \text{ , }\epsilon =\varepsilon -\frac{i\hbar \Omega }{\sqrt{\alpha
_{1}\alpha _{2}}}
\label{eqt11}
\end{align}
Now, in order to solve \ref{eqt10}, we use the following change in the variable $p$:
\begin{equation}
p\longrightarrow \rho =\frac{1}{k}\arcsin \left( \sqrt{\alpha _{2}}p\right) \text{ and } k=\hbar \sqrt{\alpha _{1}\gamma\gamma ^{\ast }}
\label{eqt12}
\end{equation}
So the equation becomes ($\delta =-i\hbar \Omega /k\alpha _{1}\alpha _{2}$):
\begin{equation}
\left\{ \frac{\partial ^{2}}{\partial \rho ^{2}}+2\delta \tan \left( k\rho\right) \frac{\partial }{\partial \rho }-\eta \tan ^{2}\left( k\rho \right)+\epsilon \right\} \psi \left( \rho \right) =0
\label{eqt13}
\end{equation}

We use the following transformation:
\begin{equation}
\psi \left( \rho \right) =\left( 1-u^{2}\right) ^{\frac{1}{2}\left( \nu +\frac{\delta }{k}\right) }f\left( u\right)
\label{eqt14}
\end{equation}
where $\nu $\ is a constant to be determined later and $u=\sin \left( k\rho\right) $. By means of this substitution \ref{eqt14}, the differential equation for $f\left( u\right) $ reduces to the following form:
\begin{equation}
\left[ \left( 1-u^{2}\right) \frac{\partial ^{2}}{\partial u^{2}}-\left(
2\nu +1\right) u\frac{\partial }{\partial u}+\frac{\epsilon }{k^{2}}-\left(
\nu +\frac{\delta }{k}\right) \right] f\left( u\right) =0  \label{eqt15}
\end{equation}
where we have chosen:
\begin{equation}
\nu \left( \nu -1\right) =\frac{\delta }{k}\left( \frac{\delta }{k}+1\right)
+\frac{\eta }{k^{2}}  \label{eqt16}
\end{equation}
In order to avoid complex eigenvalues of $\epsilon $, we must impose the condition $\Omega =0$ to eliminate the imaginary part in \ref{eqt11}; this fixes the value of the arbitrary parameter $\lambda $:
\begin{equation}
\lambda =\frac{1}{\gamma \gamma ^{\ast }}=\frac{\alpha _{1}}{m^{2}\omega
^{2}\alpha _{2}+\alpha _{1}}  \label{eqt17}
\end{equation}
This also brings the differential equation \ref{eqt15} to the Gegenbauer form \cite{Gradshteyn}:
\begin{equation}
\left[ \left( 1-u^{2}\right) \frac{\partial ^{2}}{\partial u^{2}}-\left(
2\nu+1\right) u\frac{\partial }{\partial u}+n\left( n+2\nu\right) \right]
f\left( u\right) =0  \label{eqt18}
\end{equation}
where $n$ and $\nu$ satisfy the relations:
\begin{equation}
\frac{\varepsilon }{k^{2}}-\nu =n\left( n+2\nu\right) \text{ and }\nu \left(
\nu -1\right) =\frac{\eta }{k^{2}}=\frac{1}{k^{2}}\left( \frac{m^{2}\omega
^{2}}{m^{2}\omega ^{2}\alpha _{2}+\alpha _{1}}-m\hbar \omega \right)
\label{eqt19}
\end{equation}
Solving the second relation for $\nu$ gives:
\begin{equation}
\nu =\frac{1}{2}\left( 1\pm \sqrt{1+\frac{4}{k^{2}}\left( \frac{m^{2}\omega
^{2}}{m^{2}\omega ^{2}\alpha _{2}+\alpha _{1}}-m\omega \hbar \right) }\right)
\label{eqt20}
\end{equation}
From expression \ref{eqt14}, we see that $f\left( u\right) $ should be non-singular at $u=\pm 1$, so the right value of $\nu $ is:
\begin{equation}
\nu =\frac{1}{2}\left( 1+\sqrt{1+\frac{4}{k^{2}}\left( \frac{m^{2}\omega ^{2}}{m^{2}\omega ^{2}\alpha _{2}+\alpha _{1}}-m\omega \hbar \right) }\right)
\label{eqt21}
\end{equation}
Now, the solution of \ref{eqt18} can be expressed in terms of Gegenbauer polynomials:
\begin{equation}
f\left( u\right) =\Lambda C_{n}^{\nu}\left( u\right)  \label{eqt22}
\end{equation}
Consequently, the expression of the wave function $\psi \left( \rho \right) $ is:

\begin{equation}
\psi _{n}\left( \rho \right) =\Lambda \left( 1-u^{2}\right)
^{\nu/2}C_{n}^{\nu}\left( u\right) =\Lambda \left( 1-\alpha _{2}p^{2}\right)
^{\nu/2}C_{n}^{\nu}\left( \sqrt{\alpha _{2}}p\right)  \label{eqt23}
\end{equation}

We can obtain the normalization constant $\Lambda $ by applying the deformed
normalization condition:
\begin{equation}
\int_{-\infty }^{+\infty }\frac{dp}{\left( 1-\alpha _{2}p^{2}\right) ^{1/2}}\psi ^{\ast }\left( p\right) \psi \left( p\right) =1  \label{eqt24}
\end{equation}
Then using the following identity \cite{Gradshteyn}:
\begin{equation}
\int_{-1}^{+1}dq\left( 1-q^{2}\right) ^{\nu-\frac{1}{2}}\left[ C_{n}^{\nu}\left(q\right) \right] ^{2}=\frac{\pi 2^{1-2\nu}\Gamma \left( 2\nu+n\right) }{n!\left(n+\nu\right) \left[ \Gamma \left( \nu\right) \right] ^{2}}
\label{eqt25}
\end{equation}
we get:
\begin{equation}
\Lambda =\frac{2^{\nu}\alpha _{2}^{1/4}}{\sqrt{2\pi }}\sqrt{\frac{n!\left(n+\nu\right) \left[ \Gamma \left( \nu\right) \right] ^{2}}{\Gamma \left(2\nu+n\right) }}  \label{eqt26}
\end{equation}
This completes the determination of the wave functions \ref{eqt23}.

Let us now check these solutions by studying the limits $\alpha_{1}\rightarrow 0$ and $\alpha _{2}\rightarrow 0$ from \ref{eqt21}. Using the following relations \cite{Gradshteyn}:
\begin{equation}
\lim_{\nu \rightarrow \infty }\nu ^{-\frac{n}{2}}C_{n}^{\frac{\nu}{2}}\left( x\sqrt{\frac{2}{\nu }}\right) =\frac{1}{\sqrt{2^{n}}n!}H_{n}\left( x\right)\text{ , }\lim_{\nu \rightarrow \infty }\frac{\Gamma \left( \nu+a\right) }{\Gamma \left( \nu\right) }\nu ^{-a}=1
\label{eqt27}
\end{equation}
with the doubling formula:
\begin{equation}
\Gamma \left( 2\nu\right) =\frac{2^{2\nu -1}}{\sqrt{\pi }}\Gamma \left(\nu\right) \Gamma \left( \nu+\frac{1}{2}\right)   \label{eqt28}
\end{equation}
and limiting ourselves to the first order of $\alpha _{2}$ ($\left( 1-\alpha_{2}p^{2}\right) ^{\nu/2}=\exp \left( -p^{2}/2m\omega \hbar \right) $), we obtain directly the momentum space eigenfunction of the usual KGO (without deformation):
\begin{equation}
\psi _{n}\left( p\right) =\frac{1}{\sqrt{2^{n}}n!}\left( \frac{1}{\pi m\omega \hbar }\right) ^{1/4}\exp \left( -\frac{p^{2}}{2m\omega \hbar }\right) H_{n}\left( \frac{p}{\sqrt{m\omega \hbar }}\right)
\label{eqt29}
\end{equation}
Now, to seek the energies, we employ equation \ref{eqt20} in the first relation of \ref{eqt18} and the expression of $\varepsilon $. It is straightforward
to show that the deformed KGO energy spectrum $E_{n}$ is:
\begin{equation}
E_{n}=\pm mc^{2}\sqrt{1+\frac{2\omega \hbar }{mc^{2}}n+\frac{\hbar ^{2}\left( m^{2}\omega ^{2}\alpha _{2}+\alpha _{1}\right) }{m^{2}c^{2}}n^{2}}\text{ , }n\in \mathbb{N}
\label{eqt30}
\end{equation}
Due to the modification of the Heisenberg algebra, these expressions of the energy spectrum contains an additional correction term coming from the deformation and its values increases with both parameters $\alpha _{1}$ and $\alpha_{2}$. Here it should be noted that, according to the $n^{2}$ dependence of the energy levels which explains confinement at the high energy area, our result is similar to the energies of a spinless relativistic quantum particle moving in a square well potential whose boundaries are placed at $\pm \pi /2\sqrt{m^{2}\omega ^{2}\alpha _{2}+\alpha_{1}}$.

The shape of our energy spectrum can be tested as follows: Using the limit $\alpha _{1}\rightarrow 0$, we obtain the same results as in the 1-dim KGO case with the presence of minimal length \cite{Chargui} (with $\beta =\alpha_{2}$ and $\beta ^{\backprime }=0$). Subsequently, when the deformed parameters are absent, i.e. $\alpha _{1}=\alpha _{2}=0$, the result is
strictly consistent with the KGO in 1-dim \cite{Xiao}.

\begin{figure}[tbp]
\centering
\includegraphics[width=0.5\textwidth]{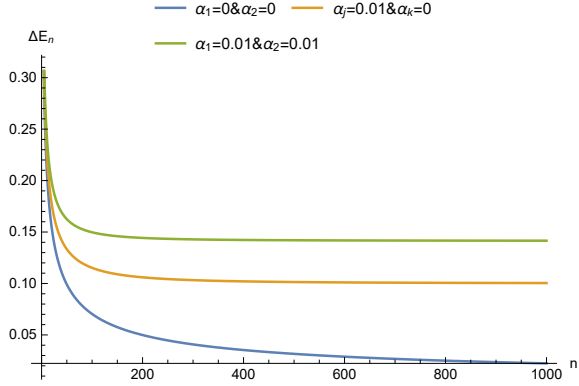}
\caption{Spectrum spacing $\Delta E_{n}$ with and without deformations}
\label{fig1}
\end{figure}

Another interesting characteristic appears in our results when computing the energy levels spacing; this difference becomes constant for large values of $n$:
\begin{equation}
\lim_{n\rightarrow \infty }\left\vert \Delta E_{n}\right\vert =\hbar c\sqrt{m^{2}\omega ^{2}\alpha _{2}+\alpha _{1}}=\hbar\omega mc^{2}\sqrt{\theta }\text{ with }\theta =\frac{1}{c^{2}}\left( \frac{\alpha _{1}}{m^{2}\omega
^{2}}+\alpha _{2}\right)  \label{eqt31}
\end{equation}
In order to show this behavior, we have plotted the energy levels spacing versus the quantum number $n$ for different values of the deformation parameters in figure \ref{fig1} (we use the units $\hbar =c=m=1$ and we put $\omega =1$).

We see from this figure \ref{fig1} that when $\alpha _{1}$ and $\alpha _{2}$ tends to $0$, that the spacing $\Delta E_{n}$ between energy levels tends to zero for large values of $n$. i.e. the energy spectrum becomes continuous at this limit in the ordinary case (without deformation). By cons, this continuous feature of the spectrum disappears and reduces to a bounded spectrum in the deformed case with a spacing value proportional to the deformation parameter $\theta $. The smallness of this parameter $\theta $ makes that the spectrum appear almost continuous in the deformed case and this also explains its continuous nature in the ordinary case.

Now, to obtain an upper bound on the parameters of deformation $\alpha _{1}$ and $\alpha _{2}$; we use \ref{eqt30} and we expand up to the first order in $\theta $:
\begin{equation}
E_{n}=mc^{2}\sqrt{1+\frac{2\omega \hbar n}{mc^{2}}}+\frac{\hbar ^{2}\omega
^{2}mc^{2}n^{2}}{2\sqrt{1+\frac{2\omega \hbar n}{mc^{2}}}}\theta
\label{eqt32}
\end{equation}
The deviation of the $n$-th energy level from the usual case caused by the modified commutation relations \ref{eqt1} is given by:
\begin{equation}
\frac{\Delta E_{n}}{\hbar \omega }=\frac{\hbar \omega mc^{2}n^{2}}{2\sqrt{1+\frac{2\omega \hbar n}{mc^{2}}}}\theta  \label{eqt33}
\end{equation}
We use the experimental results of the cyclotron motion of an electron in a Penning trap \cite{Brown}. In the absence of deformations, the cyclotron frequency of an electron of mass $m_{e}$ trapped in a magnetic field of strength $B$ is $\omega _{c}=eB/m_{e}$, therefore for a magnetic field of strength $B=6T$ we have $m_{e}\hbar \omega _{c}=e\hbar B=10^{-52} kg^{2}m^{2}s^{-2}$. At this stage, if we assume that at the level $n=10^{10}$, only deviations of the scale of $\hbar \omega _{c}$ can be detected and by taking $\Delta E_{n}<\hbar \omega _{c}$ (no perturbation of the $n$-th energy level is observed) \cite{Chang}, we can write the following constraint:
\begin{equation}
\theta <10^{33}c^{-2}kg^{-2}m^{-2}s^{2}  \label{eqt34}
\end{equation}
This leads to the following upper bounds according to the parameters of deformation $\alpha _{1}$ and $\alpha _{2}$; for $\alpha _{1}\neq 0$ and $ \alpha _{2}=0$ we get:
\begin{equation}
\Delta X_{\min }=\hbar \sqrt{\alpha _{2}}<3.33\times 10^{-18}m  \label{eqt35}
\end{equation}
and for $\alpha _{1}=0$ and $\alpha _{2}\neq 0$ we have:
\begin{equation}
\Delta P_{\min }=\hbar \sqrt{\alpha _{1}}<3.17\times 10^{-36}kgms^{-1}
\label{eqt36}
\end{equation}
For the non-relativistic limit, by setting $E=m c^{2}+E_{nr}$ with the assumption that $m c^{2}\gg E_{nr}$, we write the spectrum of the non-relativistic KGO in the deformed SdS space:
\begin{equation}
E_{nr}=n\hbar \omega \left( 1+n\frac{\hbar }{2m\omega }\left( m^{2}\omega
^{2}\alpha _{2}+\alpha _{1}\right) \right)  \label{eqt37}
\end{equation}

\section{D-dimensional Klein-Gordon oscillator}

\label{sec:ND}

In this section, we consider the stationary relativistic equation describing the D-dimensional (D-dim) KGO in momentum representation:
\begin{equation}
\left[ c^{2}\left( \mathbf{P}+im\omega \mathbf{X}\right) \left( \mathbf{P-}im\omega \mathbf{X}\right) +m^{2}c^{4}-E^{2}\right] \psi \left( \mathbf{P}\right) =0  \label{eqt38}
\end{equation}
which can be written with the help of commutation relation \ref{eqt1}$\ $as:
\begin{equation}
\left[ \left( m^{2}\omega ^{2}-m\omega \hbar \alpha _{1}\right) \mathbf{X}^{2}+\left( 1\mathbf{-}m\omega \hbar \alpha _{2}\right) \mathbf{P}^{2}-m\omega \hbar \sqrt{\alpha _{1}\alpha _{2}}\left( \mathbf{PX}+\mathbf{XP}\right) -\varepsilon ^{\backprime }\right] \psi \left( \mathbf{P}\right) =0
\label{eqt39}
\end{equation}
where:
\begin{equation}
\varepsilon ^{\backprime }=\frac{E^{2}-m^{2}c^{4}}{c^{2}}+Dm\omega \hbar
\label{eqt40}
\end{equation}
We use the following separation of the wave function into angular and radial parts :
\begin{equation}
\psi \left( \mathbf{P}\right) =Y_{l_{\left( D-1\right)}.....l_{2}l_{1}}^{m_{l}}\left( \mathbf{P}\right) \varphi \left( p\right)
\label{eqt41}
\end{equation}
where $Y_{l_{\left( D-1\right) }.....l_{2}l_{1}}^{m_{l}}\left( \mathbf{P}\right) $ are D-dim ultra-spherical harmonics; this enables us to make the following replacements in the momentum space ($l$ is the orbital quantum number):
\begin{equation}
\sum\limits_{i=1}^{D}\frac{\partial ^{2}}{\partial p_{i}^{2}}=\frac{\partial
^{2}}{\partial p^{2}}+\frac{D-1}{p}\frac{\partial }{\partial p}-\frac{L^{2}}{p^{2}}\text{ , }\sum\limits_{i=1}^{D}p_{i}\frac{\partial }{\partial p_{i}}=p\frac{\partial }{\partial p}\text{ , }L^{2}=l\left( l+D-2\right)
\label{eqt42}
\end{equation}
Using the definition of the momentum space realization of the operators $\mathbf{X}$ and $\mathbf{P}$ in \ref{eqt6} and \ref{eqt7}, we obtain the following equation in the deformed momentum space:
\begin{equation}
\left\{
\begin{array}{c}
\hbar ^{2}\frac{\alpha _{1}}{\alpha _{2}}\gamma \gamma ^{\ast }\left[ \left(
\sqrt{1-\alpha _{2}p^{2}}\frac{\partial }{\partial p}\right) ^{2}+\left(
1-\alpha _{2}p^{2}\right) \left( \frac{D-1}{p}\frac{\partial }{\partial p}-
\frac{L^{2}}{p^{2}}\right) \right] \\
-\frac{2i\hbar \Omega }{\sqrt{\alpha _{1}\alpha _{2}}}p\frac{\partial }{
\partial p}-\frac{\eta \alpha _{2}p^{2}}{1-\alpha _{2}p^{2}}+\epsilon
^{\backprime }
\end{array}
\right\} \varphi \left( p\right) =0  \label{eqt43}
\end{equation}
where, the expressions of $\gamma $, $\Omega $ and $\eta $ are given in \ref{eqt11} and $\epsilon ^{\backprime }$ is defined by:
\begin{equation}
\epsilon ^{\backprime }=\varepsilon ^{\backprime }-\frac{Di\hbar \Omega }{\sqrt{\alpha _{1}\alpha _{2}}}  \label{eqt44}
\end{equation}
Now, in order to solve \ref{eqt43}, we use the same transformation \ref{eqt12} of the 1-dim case to get:
\begin{equation}
\left\{
\begin{array}{c}
\frac{\partial ^{2}}{\partial \rho ^{2}}+\left( k\left( D-1\right) \cot
\left( k\rho \right) -\frac{2i\hbar \Omega }{k\sqrt{\alpha _{1}\alpha _{2}}}
\tan \left( k\rho \right) \right) \frac{\partial }{\partial \rho } \\
-k^{2}l\left( l+D-2\right) \cot ^{2}\left( k\rho \right) -\eta \tan
^{2}\left( k\rho \right) +\epsilon ^{\backprime }
\end{array}
\right\} \varphi \left( \rho \right) =0  \label{eqt45}
\end{equation}
At this point, in order to avoid complex eigenvalues of the energies, we must also impose the same condition as in 1-dim case \ref{eqt17} to eliminate the imaginary term; this transforms the master equation \ref{eqt45} to the following:
\begin{equation}
\left\{
\begin{array}{c}
\frac{\partial ^{2}}{\partial \rho ^{2}}+k\left( D-1\right) \cot \left(
k\rho \right) \frac{\partial }{\partial \rho }-k^{2}l\left( l+D-2\right)
\cot ^{2}\left( k\rho \right) \\
-\left( \frac{1-\lambda }{\alpha _{2}}-m\omega \hbar \right) \tan ^{2}\left(
k\rho \right) +\varepsilon ^{\backprime }
\end{array}
\right\} \varphi \left( \rho \right) =0  \label{eqt46}
\end{equation}
In order to simplify this equation, we use the following ansatz:
\begin{equation}
\varphi \left( \rho \right) =\left( 1-q^{2}\right) ^{\mu /2}q^{l}f\left(q\right)  \label{eqt47}
\end{equation}
where $\mu $ is a constant to be determined and $q=sin\left( k\rho \right) $.

Doing so, the differential equation for $f\left( q\right) $ \ref{eqt46} is written as:
\begin{equation}
\left\{ \left( 1-q^{2}\right) \frac{\partial ^{2}}{\partial q^{2}}+\left[
-\left( 2\mu +2l+D\right) q+\frac{2l+D-1}{q}\right] \frac{\partial }{\partial q} +\frac{\varepsilon ^{\backprime }}{k^{2}}-\left( 2l-D\right) \mu-l\right\} f\left( q\right) =0
\label{eqt48}
\end{equation}
where $\mu $ is chosen to simplify this equation as:
\begin{equation}
\mu (\mu -1)-\frac{1}{k^{2}}\left( \frac{1-\lambda }{\alpha _{2}}-m\omega
\hbar \right) =0  \label{eqt49}
\end{equation}
From expression \ref{eqt47}, we see that $f\left( q\right) $ should be non-singular when $q=\pm 1$, which implies:
\begin{equation}
\mu =\frac{1}{2}\left( 1+\sqrt{1+\frac{4}{k^{2}}\left( \frac{m^{2}\omega ^{2}%
}{m^{2}\omega ^{2}\alpha _{2}+\alpha _{1}}-m\omega \hbar \right) }\right)
\label{eqt50}
\end{equation}

We note that equation \ref{eqt48} possesses three singular points $q=0,1,-1$. In order to reduce this equation to a class of known differential
equations with a polynomial solution, we use another change in the variable;
we write $z=2q^{2}-1$ and impose the following condition:%
\begin{equation}
n_{r}\left(n_{r}+a+b+1\right)=\frac{1}{4}\left[\frac{\varepsilon^{\backprime}}{k^{2}}-\left(2l-D\right)\mu-l\right]  \label{eqt51}
\end{equation}%
where $n_{r}$ is a non-negative integer (radial quantum number) and we have also defined:
\begin{equation}
a=\mu -1/2\text{ , }b=l-1+D/2
\label{eqt52}
\end{equation}
Now, the differential equation \ref{eqt48} transforms to the following Jacobi form:
\begin{equation}
\left(1-z^{2}\right)\frac{d^{2}g}{dz^{2}}+\left[\left(b-a\right)-\left(a+b+2\right)z\right]\frac{dg}{dz} +n_{r}\left(n_{r}+1+a+b\right)g\left(z\right)=0
\label{eqt53}
\end{equation}
The solutions are given by the Jacobi polynomials $g(z)=P_{n_{r}}^{\left(a,b\right)}\left(z\right)$ and so the radial wave function $\varphi_{n_{r},l}\left( p\right)$ is:
\begin{equation}
\varphi_{n_{r},l}\left(p\right) =N\left(1-\alpha_{2}p^{2}\right)^{\mu/2}\left(\alpha_{2}p^{2}\right)^{l/2}P_{n_{r}}^{\left(a,b\right)}\left(2\alpha_{2}p^{2}-1\right)
\label{eqt54}
\end{equation}

To determine the normalization constant $N$, we use the deformed normalization condition in the D-dim space of radial wave functions:
\begin{equation}
\int_{0}^{1/\sqrt{\alpha_{2}}}\frac{Dp^{D-1}dp}{\left(1-\alpha_{2}p^{2}\right)^{1/2}}\psi^{\ast}\left(p\right)\psi\left(p\right)=1
\label{eqt55}
\end{equation}
and the Jacobi polynomial orthogonality relation \cite{Gradshteyn}:
\begin{equation}
\int_{-1}^{1}dy\left(1-y\right)^{\alpha}\left(1+y\right)^{\beta}\left[P_{n_{r}}^{\left(\alpha,\beta\right)}\left(y\right)\right]^{2} =\frac{\pi 2^{\alpha+\beta+1}\Gamma \left(\alpha +n_{r}+1\right)\Gamma\left(\beta+n_{r}+1\right)} {n_{r}!\left(\alpha+\beta+1+2n_{r}\right)\Gamma\left(\alpha+\beta+n_{r}+1\right)}
\label{eqt56}
\end{equation}
so we get the expression:
\begin{equation}
N=\frac{2\alpha_{2}^{D/4}}{\sqrt{2\pi}}\left[\frac{n_{r}!\left(2n_{r}+\mu+l+\frac{D-1}{2}\right)\Gamma\left(n_{r}+\mu+l +\frac{D-1}{2}\right)}{D\Gamma\left(n_{r}+\mu +\frac{1}{2}\right)\Gamma\left(n_{r}+l+\frac{D}{2}\right)}\right]^{1/2}
\label{eqt57}
\end{equation}
Now to find the energies, we substitute the relations \ref{eqt40} and \ref{eqt50} into the condition \ref{eqt51} and we define the principal quantum number $n$ by $n=2n_{r}+l$, so we obtain the spectrum of the D-dim SdS KGO as follows:
\begin{equation}
E_{n,l}=\pm mc^{2}\left\{ 1+\frac{2\hbar \omega }{mc^{2}}n+\frac{\hbar
^{2}\left( \alpha _{1}+m^{2}\omega ^{2}\alpha _{2}\right) }{m^{2}c^{2}}\left[
n^{2}+\left( D-1\right) n-l\left( l+D-2\right) \right] \right\} ^{1/2}
\label{eqt58}
\end{equation}
The expression shows that we have the same term as in the 1-dim case with a small offset and a new correction containing the orbital quantum number $l$; It lifts the degeneracy of the energies which remains, however $(2l+1)$.
The relation \ref{eqt58} also shows that the contributions coming from the deformations increase with the dimension of the space $D$ but decrease with the quantum number $l$. This shape of the energy spectrum can be tested by taking the limit $\alpha _{1}\rightarrow 0$ and $\alpha _{2}\rightarrow 0$ where we obtain the exact result of the D-dim KGO without deformation \cite
{Chargui, Garcia}. It also gives to same spectrum of the two-dimensional case obtained in \cite{Falek19} (when $B=0$ in this later).

We also mention that the energy levels spacing $\Delta E_{n}$ for large values of $n$ in this case is given with the same limit of the 1-dim \ref{eqt31}.

\section{Thermal Properties}

\label{sec:TP}

In order to obtain the thermodynamic properties of the deformed D-dim KGO with SdS commutation relations, we consider the partition function of the system at finite temperature $T$ according to the Boltzmann factor as follows:
\begin{equation}
Z=\sum_{n=0}^{\infty }e^{-\frac{E_{n,l}}{k_{B}T}}=\sum_{n=0}^{\infty }\exp
\left( -\frac{mc^{2}}{k_{B}T}\sqrt{a_{1}+a_{2}n+a_{3}n^{2}}\right)
\label{eqt59}
\end{equation}
where $k_{B}$\ is the Boltzmann constant and the other parameters of the expression are defined as follows:
\begin{equation}
a_{1}=1-a_{3}l\left( l+D-2\right) \text{ , }a_{2}=\frac{2\omega \hbar }{mc^{2}}+a_{3}\left( D-1\right) \text{ and }a_{3}=\frac{\hbar ^{2}\left(\alpha _{1}+m^{2}\omega ^{2}\alpha _{2}\right) }{m^{2}c^{2}}  \label{eqt60}
\end{equation}
To evaluate the function \ref{eqt59}, we use the Euler-MacLaurin formula:
\begin{equation}
\sum_{n=0}^{\infty }f\left( n\right) =\frac{1}{2}f\left( 0\right)+\int_{0}^{\infty }f\left( x\right) dx-\sum_{p=1}^{\infty }\frac{1}{\left(2p\right) !}B_{2p}f^{\left( 2p-1\right) }\left( 0\right)
\label{eqt61}
\end{equation}
where $B_{2p}$ are the Bernoulli numbers, $f^{(2p-1)}$ designates the derivative of order $\left( 2p-1\right) $. The integral in the expression is denoted $I$ and is given as follows:
\begin{equation}
I=\frac{2a_{1}}{\sqrt{a_{2}^{2}-4a_{1}a_{3}}}\sum_{n=0}^{\infty }\frac{\left( 2n-1\right) !!}{\left( 2n\right) !!}\left( \frac{-4a_{1}a_{3}}{a_{2}^{2}-4a_{1}a_{3}}\right) ^{n}\left[ \frac{\Gamma \left( 2n+2\right) }{\chi ^{2n+2}}-\frac{e^{-\chi }}{2n+2}\Phi \left( 1,2n+2,\chi \right) \right]
\label{eqt62}
\end{equation}
where $\chi=\frac{mc^{2}}{k_{B}T}\sqrt{a_{1}}$ and we have used the new variable $y=\sqrt{1+\frac{a_{2}}{a_{1}}n+\frac{\alpha _{3}}{a_{1}}n^{2}}$ and also the power series of the square root of the following integral:
\begin{equation}
I^{\backprime}=\frac{2a_{1}}{\sqrt{a_{2}^{2}-4a_{1}a_{3}}}\int_{1}^{\infty }dyy\left( 1+\frac{4a_{1}a_{3}}{a_{2}^{2}-4a_{1}a_{3}}y^{2}\right) ^{-1/2}e^{-\chi y}
\label{eqt63}
\end{equation}
At high temperatures, we note that the contributions of the first and the third terms in expression \ref{eqt61} and that of the second term in \ref{eqt62} become negligible compared to the term in $\chi ^{-\left(2n+2\right) }$. Therefore the partition function can be written as:
\begin{equation}
Z\simeq \frac{2\left( \frac{k_{B}T}{mc^{2}}\right) ^{2}}{\sqrt{a_{2}^{2}-4a_{1}a_{3}}}\sum_{n=0}^{\infty }\left( -1\right) ^{n}\Gamma\left( 2n+2\right) \frac{\left( 2n-1\right) !!}{\left( 2n\right) !!}\sigma^{n}  \label{eqt64}
\end{equation}
with:
\begin{equation}
\sigma =\left( \frac{k_{B}T}{mc^{2}}\right) ^{2}\left( \frac{4a_{3}}{a_{2}^{2}-4a_{1}a_{3}}\right)  \label{eqt65}
\end{equation}
Keeping only the orders 0 and 1 contributions in both $\alpha _{1}$ and $\alpha _{2}$, we get the simplified form:
\begin{equation}
Z\simeq \frac{\left( k_{B}T\right) ^{2}}{m\omega \hbar c^{2}}\left[ 1-\left(
\frac{\alpha _{1}}{m^{2}\omega ^{2}}+\alpha _{2}\right) \left[ \frac{3\left(
k_{B}T\right) ^{2}}{c^{2}}\left( 1-\frac{1}{6}\left( \frac{mc^{2}}{k_{B}T}
\right) ^{2}\right) +\frac{\left( D-1\right) \hbar }{2m\omega }\right]
\right]  \label{eqt66}
\end{equation}
Considering that $k_{B}T\gg mc^{2}$ in the case of high temperatures, we finally obtain the expression of the partition function in this high-temperature regime:
\begin{equation}
Z\simeq \frac{\left( k_{B}T\right) ^{2}}{m\omega \hbar c^{2}}\left[ 1-\theta
\left( 3\left( k_{B}T\right) ^{2}+\frac{\left( D-1\right) \hbar c^{2}}{
2m\omega }\right) \right]  \label{eqt67}
\end{equation}
where the expression of $\theta $ is the same as in 1-dim case \ref{eqt31}.

According to their definitions, we can obtain the thermodynamic properties of our system, such as the free energy $F$, the mean energy $U$, the specific heat $C$ and the entropy $S$, as follows:
\begin{equation}
F=-k_{B}T\ln Z=-k_{B}T\ln \left[ \frac{\left( k_{B}T\right) ^{2}}{m\omega
\hbar c^{2}}\left( 1-\theta \delta \right) \right]   \label{eqt68}
\end{equation}
\begin{equation}
U=k_{B}T^{2}\frac{\partial \ln Z}{\partial T}=4k_{B}T\left[ 1-\frac{2m\omega
-\hbar c^{2}\theta \left( D-1\right) }{4m\omega -2\theta \left( 3m\omega
\left( k_{B}T\right) ^{2}+m\omega \delta \right) }\right]   \label{eqt69}
\end{equation}
\begin{equation}
C=\frac{\partial U}{\partial T}=4k_{B}\left[ 1-\frac{\left( 2m\omega -\hbar
c^{2}\theta \left( D-1\right) \right) \left( 4m\omega +2\theta \left(
9m\omega \left( k_{B}T\right) ^{2}-m\omega \delta \right) \right) }{\left(
4m\omega -2\theta \left( 3m\omega \left( k_{B}T\right) ^{2}+m\omega \delta
\right) \right) ^{2}}\right]   \label{eqt70}
\end{equation}
\begin{equation}
S=-\frac{\partial F}{\partial T}=k_{B}\left[ \frac{2\left( 1-\theta \left(
3\left( k_{B}T\right) ^{2}+\delta \right) \right) }{\left( 1-\theta \delta
\right) }+\ln \left[ \frac{\left( k_{B}T\right) ^{2}}{m\omega \hbar c^{2}}
\left( 1-\theta \delta \right) \right] \right]   \label{eqt71}
\end{equation}
where $\delta =3\left( k_{B}T\right) ^{2}+\frac{1}{2}\left( D-1\right) \hbar
c^{2}m^{-1}\omega ^{-1}$.

\begin{figure}[tbp]
\centering
\includegraphics[width=0.5\textwidth]{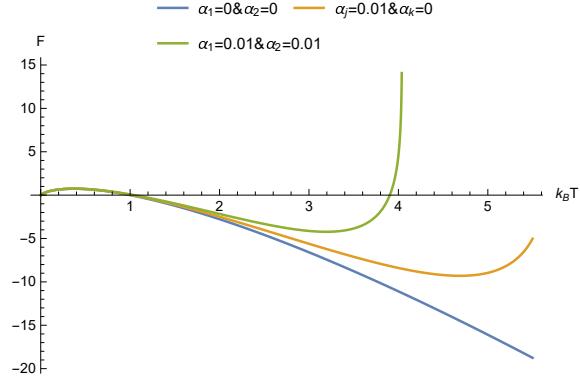}
\caption{Effects of the deformations on the free energy $F$}
\label{Fig2}
\end{figure}

\begin{figure}[tbp]
\centering
\includegraphics[width=0.5\textwidth]{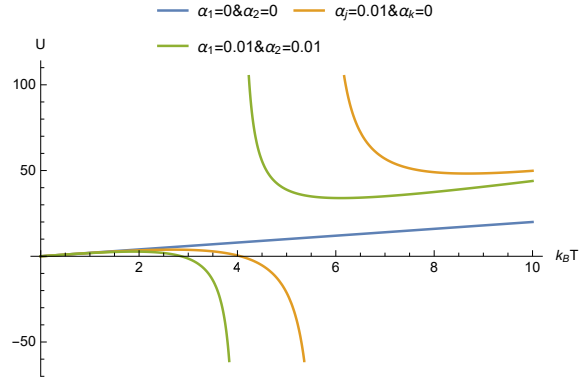}
\caption{Effects of the deformations on the energy $U$}
\label{Fig3}
\end{figure}

\begin{figure}[tbp]
\centering
\includegraphics[width=0.5\textwidth]{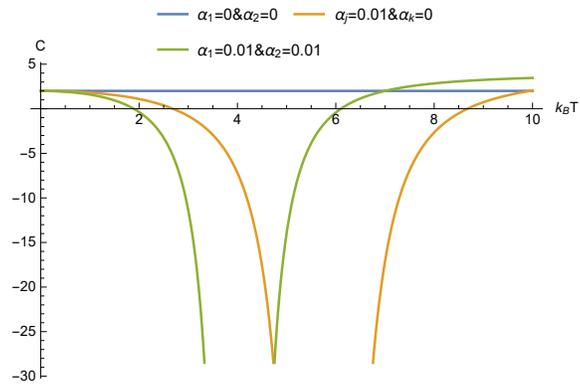}
\caption{Effects of the deformations on the heat capacity $C$}
\label{Fig4}
\end{figure}

\begin{figure}[tbp]
\centering
\includegraphics[width=0.5\textwidth]{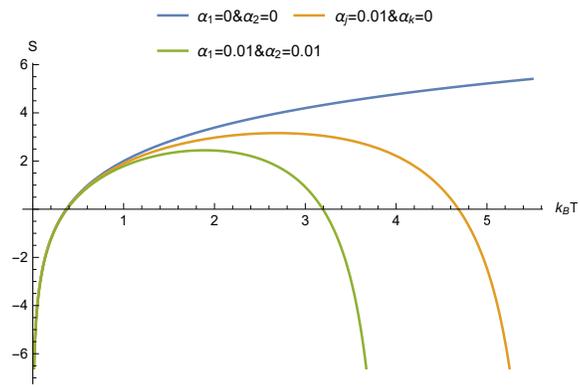}
\caption{Effects of the deformations on the entropy $S$}
\label{Fig5}
\end{figure}

We show the dependence of these thermodynamic properties with the temperature in the 3D case for a few indicative values of the parameters of deformations in figures \ref{Fig2}, \ref{Fig3}, \ref{Fig4} and \ref{Fig5}. We see in these plots that the corrections increase the free energy $F$ but decrease the other thermodynamic properties. We mention here that the space dimension does not affect the shape of these curves but it just shifts them towards low temperatures.

The expressions \ref{eqt67} to \ref{eqt71} give the thermodynamic properties of the usual D-dim KGO when using the limit $\theta \rightarrow 0$ (or equivalently for $\alpha _{1}\rightarrow 0$ and $\alpha _{2}\rightarrow 0$).

\section{Conclusion}
\label{sec:CC}

We studied, with an explicit calculation in the momentum space representation, the Klein-Gordon oscillator in arbitrary dimension in the framework of deformed quantum mechanics with Snyder-de Sitter commutation relations; these deformations leads to a non zero minimal uncertainty in the measurement of both position and momentum of the spinless particle.

In one dimensional space, the exact expression of the energy levels is obtained and the wave function is expressed with the Gegenbauer polynomials. The associated energy spectrum is found with an additional correction to the usual one; this correction depends on the deformation parameters of both Snyder algebra and de Sitter space. This corrections grow quickly with $n$ and this can be associated with an additional confinement to the usual one representing the harmonic oscillator. Within large values of $n$, the corrections cause the spectrum to tend towards a discrete one with a spacing proportional to both parameters of the deformations; this explains the fact that the spectrum of the ordinary case (without deformations) is almost continuous in this limit.

For the case of arbitrary dimension, we obtain the normalized momentum space wave function in term of the Jacobi polynomials. Concerning the associated energy levels, we found an additional term depending on the orbital quantum number $l$ which was absent in the non-deformed case. Therefore, the contributions of Snyder algebra and the effects of gravity lead to a lifting of the spectrum degeneracy found in ordinary harmonic oscillator. The energy spacing for large values of $n$ was found similar to the one dimensional case.

Regarding the thermodynamic properties of our system, we showed that in the regime of high temperatures, they have been also affected by the deformation parameters. Comparing their values with those of the ordinary case, we found that all the thermodynamic properties decrease except for the free energy $F$ which increases.

\section*{Acknowledgment}

This work was done with funding from the DGRSDT of the Ministry of Higher Education and Scientific Research in Algeria as part of the PRFU B00L02UN070120190003.

\section*{Data Availability Statement}
The data that support the findings of this study are available from the corresponding author upon reasonable request.

\end{document}